# Valley Piezoelectric Mechanism for Interpreting and Optimizing Piezoelectricity in Quantum Materials via Anomalous Hall Effect

Yilimiranmu Rouzhahong, Chao Liang, Chong Li, Biao Wang*, and Huashan Li*

*School of Physics, Sun Yat-Sen University, Guangzhou, 510275 China.*

Quantum materials have exhibited attractive electro-mechanical responses, but their piezoelectric coefficients are far from satisfactory due to the lack of fundamental mechanisms to benefit from the quantum effects. We discovered the valley piezoelectric mechanism that is absent in traditional piezoelectric theory yet promising to overcome this challenge. A theoretical model was developed to elucidate the valley piezoelectricity as the Valley Hall effect driven by pseudoelectric field, which can be significant in quantum systems with broken time reversal symmetry. Consistent tight-binding and density-functional-theory (DFT) calculations validate the model and unveil the crucial dependence of valley piezoelectricity on valley splitting, hybridization energy, bandgap, and Poisson ratio. Doping, passivation, and external stress are proposed as rational strategies to optimize piezoelectricity, with a more than 130% increase of piezoelectricity demonstrated by DFT simulations. The general valley piezoelectric model bridges the gap between electro-mechanical response and quantum effects, which opens an opportunity to achieve outstanding piezoelectricity in quantum materials via optimizing spin-valley and spin-orbit couplings.

## 1. Introduction

Given its valuable capabilities to drive fast state transition and to achieve energy conversion, piezoelectricity has been serving as the fundamental mechanism for a wide range of emerging technologies from robotics to artificial intelligence and from energy harvesters to wearable electronics.[1] Traditional bulk piezoelectric materials such as PZT and Ba/Pb-based perovskites[2] are hard, brittle, and lead-containing fail to meet the modern requirements of miniaturization, flexibility, ultrafast response, and sustainable development. Recently, the discovery of a large family of quantum materials not only refreshes our understanding of piezoelectric mechanism, but also unveils the potential to overcome the above challenges via collective quantum effects.[3] As representative quantum materials, two-dimensional (2D) transition metal dichalcogenides (TMD),[4] hexagonal BN,[5] puckered SnS[6] and a variety of predicted monolayers[4a, 7] exhibit piezoelectric response owing to symmetry breaking and quantum confinement even though none of their bulk counterparts are piezoelectric materials. Piezoelectricity has also been experimentally detected in graphene, which breaks through the restriction that piezoelectric response should be absent in semi-metallic and centrosymmetric systems.[8] Similar phenomena are computationally predicted in other 2D materials with zero bandgap such as 1H-$MX_2$ (M = Nb, Ta; X = S, Se) monolayers[9] and $Fe_2IBr$.[10]

Despite of the unprecedented phenomena, outstanding mechanical strength and high tunability achieved in quantum piezoelectric materials,[11] the fundamental mechanisms that govern the electro-mechanical response arising from quantum effects remain to be fully understood. Within the framework of modern polarization theory, a few analytical studies have been conducted to describe piezoelectricity from the quantum perspective and to successfully predict piezoelectric coefficients for BN and TMD monolayers.[12] Their results suggest that piezoelectric response crucially depends on the distribution of effective Berry curvature in reciprocal space, which can be tuned by controlling hopping energy, atomic polarization, electronegativity etc.[13] Nevertheless, the piezoelectric mechanism proposed by the above models is consistent with the general scenario that polarization is induced by wavefunction deformation, and is irrelevant to the collective quantum effects uniquely owned by quantum materials. So far, it is still unclear whether the extra degrees of freedom provided by quantum materials can be employed to manipulate the effective Berry curvature and thus the charge polarization under strain, and how the high-order quantum effects such as spin-orbit, spin-valley, and electron-phonon coupling influence such energy conversion process. These questions are crucial to design promising quantum materials, yet are impossible to be answered by existing theoretical models because of their intrinsic deficiencies: the absence of potential terms for describing atomic interactions beyond nearest neighbors; the disconnection between piezoelectricity and important quantum features such as spin and valley.

The valley degree of freedom in quantum materials has drawn great attention owing to its remarkable impacts on electrical, magnetic and optical properties.[14] Valley polarization has been observed in a variety of quantum materials including $MoS_2$, $MoSe_2$, $WS_2$, and $WSe_2$ monolayers, which can also be induced by external stimuli via breaking spatial symmetries.[15] Band splitting between spin-up and spin-down components around valley points are predicted for TMDs as a consequence of spin-orbit coupling. Based on the contrast of Berry curvatures at different valleys, Valley Hall effect has been proposed to describe polarized currents. Our previous study also found that Berry curvatures of most 2D materials exhibit mutations at valleys and rapidly decay away from the valleys.[16] Considering the dominant contributions of Berry curvature around the valley points to 2D piezoelectricity and the rich quantum phenomena associated with valley points, we speculate that optimizing the distribution and occupation of valleys in response to deformation may serve as a potential avenue for harvesting quantum effects in piezoelectric systems.

In this work, we discovered intensive valley piezoelectricity in parallel with traditional piezoelectricity in 2D quantum materials, by incorporating the scalar potential to describe next nearest neighbor interactions and the spin-orbit coupling to break the symmetry of valley distribution. The fundamental mechanism of valley piezoelectricity is elucidated to be the Valley Hall effect by our theoretical model, wherein non-vanishing Valley Hall current is activated by the pseudoelectric field originating from the scalar potential. Consistent tight-binding and first-principles calculations validate the proposed model, and suggest that the valley piezoelectricity crucially depends on valley splitting, hybridization energy, bandgap and Poisson ratio. A 130% enhancement of piezoelectricity is achieved by manipulating the valley distribution of MoS$_2$. The above finding paves an avenue to optimize piezoelectricity of quantum materials via valley piezoelectric mechanism.

## 2. Results and Discussion
### 2.1. Basic Theory of Valley Piezoelectricity

Within the framework of modern polarization theory, piezoelectricity is defined as the stress/strain induced current change in insulators with broken inversion symmetry. For quantum materials, the model for describing electron-mechanical response should encompass the traditional charge polarization arising from wavefunction deformation as well as the additional piezoelectricity induced by quantum effects.

$$e_{ijk} = (\Delta \boldsymbol{J_T} + \Delta \boldsymbol{J_Q})_i / \Delta \varepsilon_{jk} = e_{ijk}^A + e_{ijk}^V \tag{1}$$

where $\varepsilon_{jk}$ is strain tensor, $e_{ijk}$ is the total piezoelectric coefficient, $\Delta J_T$ and $\Delta J_Q$ represent the polarized currents originating from traditional charge polarization and quantum effects, respectively. In this study, we focus on the quantum effects associated with the valley degree of freedom, and employ the large family of 2D systems as representative quantum materials to illustrate the relationship between piezoelectricity and quantum effects. The additional piezoelectricity $e_{ijk}^V$ induced by valley related interactions will be referred to as valley piezoelectricity in the following discussions.

The traditional polarized current of 2D materials under strain is described as[13a]

$$\Delta \boldsymbol{J_T} = e_{ijk}^A \, \boldsymbol{A} \times \hat{Z} \tag{2}$$

where $\boldsymbol{A}$ is the vector potential induced by strain, and $\hat{Z}$ is the unit vector in out-of-plane direction. In addition to the vector potential for describing nearest neighboring interactions, external strain also generates the scalar potential that accounts for the next nearest neighboring interactions (Figure 1a).[17] The scalar potential $\Phi(\mathbf{r})$ is absent in the expression of traditional polarized current, yet crucial to understand the impact of valley on piezoelectricity. For 2D materials, the pseudoelectric field $\mathbf{E}$ deriving from $\Phi(\mathbf{r})$ can be written as

$$\mathbf{E} = -\nabla\Phi(\mathbf{r}) = -(\frac{\partial \varepsilon_{xx}}{\partial x}\vec{x} + \frac{\partial \varepsilon_{yy}}{\partial y}\vec{y}) \tag{3}$$

Consequentially, the uniaxial stress/strain in arbitrary direction would provide pseudoelectric field in both $\vec{x}$ and $\vec{y}$ directions to activate charge polarization at valleys via anomalous Valley Hall effect as discussed later.

Valley Hall effect describes the intrinsic transport property arising from inequivalent electronic structures at different valleys (Figure 1b). The Valley Hall conductivity $\sigma_{xy}^V$ of 2D materials can be calculated by the integral of Berry curvature $\Omega_{xy}(\mathbf{k})$ over the first Brillouin zone.

$$\sigma_{xy}^V = -\frac{e^2}{h}\int \frac{d^2k}{2\pi}\Omega_{xy}(\mathbf{k}) \tag{4}$$

The contribution of each valence band to the total Berry curvature is written as

$$\Omega_{k_i,\varepsilon_{jk}}^n = i\left[\langle\partial_{k_i}u_{n,k}|\partial_{\varepsilon_{jk}}u_{n,k}\rangle - \langle\partial_{\varepsilon_{jk}}u_{n,k}|\partial_{k_i}u_{n,k}\rangle\right] \tag{5}$$

where $n$ is the index of valence band, and $u_{n,k}$ is the corresponding Bloch state. For time reversal invariant system, the Valley Hall conductivity vanishes because the Berry curvature is an odd function in reciprocal space satisfying $\Omega(k) = -\Omega(-k)$. When the time reversal symmetry is broken by external magnetic field or spin-orbit coupling effect, the symmetry of Berry curvature is broken with $\Omega(k) \neq -\Omega(-k)$, leading to the valley polarization and nonzero Valley Hall conductivity (Figure 1c).

The nonzero Valley Hall conductivity can be driven by the pseudoelectric field to generate Valley Hall current ($J$) as illustrated in Figure 1d, which satisfies the relation

$$\begin{pmatrix} 0 & \sigma_{xy}^V \\ \sigma_{yx}^V & 0 \end{pmatrix}\begin{pmatrix} E_x \\ E_y \end{pmatrix} = \begin{pmatrix} J_x \\ J_y \end{pmatrix} \tag{6}$$

According to the definition of piezoelectric response, the valley piezoelectricity originating from the variation of Valley Hall current (Figure 1e) can be solved by Eq. (1-6). For 2D quantum materials, the piezoelectric coefficient tensor including both the traditional piezoelectricity and the valley piezoelectricity takes the matrix form as

$$e_{ijk} = \begin{pmatrix} e_{xx}^A + e_{xx}^V & e_{xy}^A + e_{xy}^V & 0 \\ e_{yx}^A + e_{yx}^V & e_{yy}^A + e_{yy}^V & 0 \\ 0 & 0 & 0 \end{pmatrix} \tag{7}$$

wherein all matrix elements are independent of each other in 2D materials with no symmetry. Under the restriction of isotropic Poisson ratios, the nonzero matrix elements of valley piezoelectricity in Eq. (7) satisfies the relationship $-e_{xx}^V = -e_{yx}^V = e_{xy}^V = e_{yy}^V$. For 2D materials with D$_{3h}$ symmetry, the traditional piezoelectric elements satisfy $e_{xx}^A = 0$, $e_{xy}^A = 0$, $e_{yx}^A = -e_{yy}^A$, and the total piezoelectric $e_{ijk}$ tensors can be simplified as

$$e_{ijk} = \begin{pmatrix} -e_{xx}^V & e_{xx}^V & 0 \\ -(e_{yy}^A + e_{xx}^V) & (e_{yy}^A + e_{xx}^V) & 0 \\ 0 & 0 & 0 \end{pmatrix}. \tag{8}$$

The above result suggest that the valley piezoelectric effect has the potential to strengthen piezoelectricity as well as to trigger piezoelectric response at directions that are absent in traditional scenario.

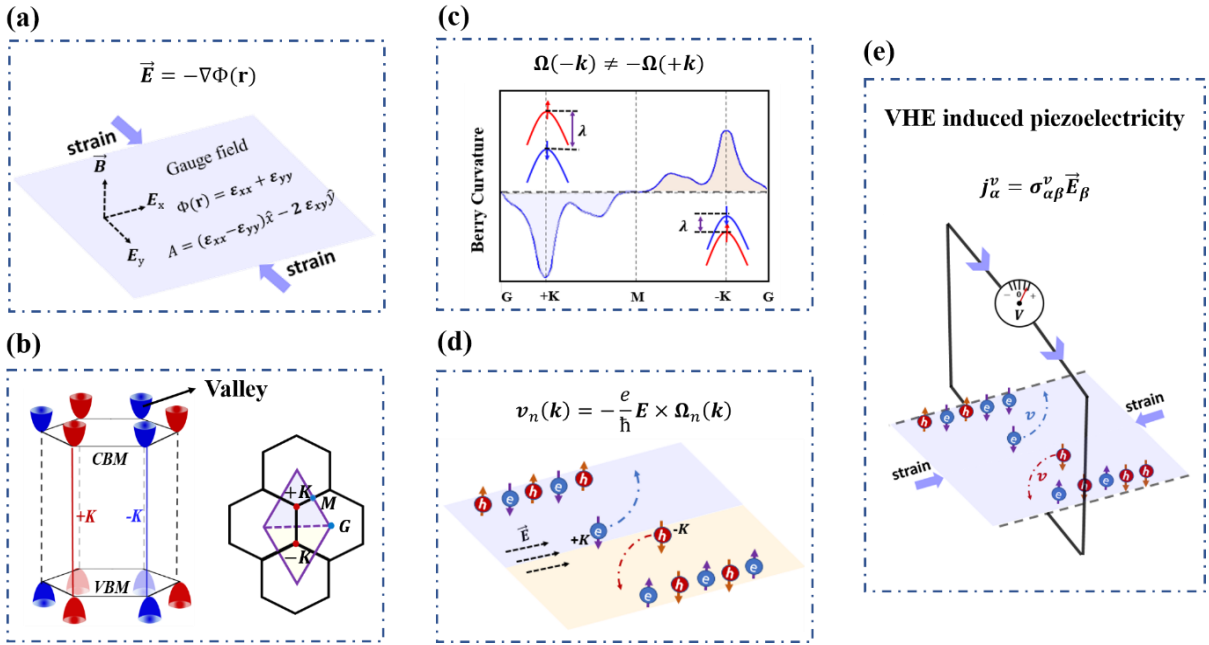

**Figure 1.** Piezoelectric mechanism originating from Valley Hall effect. (a) Illustration of strain induced gauge field. (b) Schematic plot of the first Brillouin zone and valleys for hexagonal crystal with D$_{3h}$ symmetry. Brillouin zone is partitioned into two triangular areas to reflect the opposite signs of Valley Hall conductivities. (c) Schematic illustration of asymmetric Berry curvature distribution arising from spin splitting in systems with broken time reversal symmetry. (d) Valley Hall Effect excited by strain induced pseudoelectric field and nonzero berry curvatures. (e) Valley piezoelectricity originating from Valley Hall effects.

**2.2. Dependence of Valley Piezoelectricity on Quantum Variables**

Given their large valley polarization as well as important role in electronic/spintronic applications, TMD monolayers were employed as representative systems for validating the valley piezoelectricity model via joint analytical and numerical investigations. Considering that piezoelectricity mainly depends on the electronic structure around valleys, the tight binding (TB) Hamiltonian[13c] with the consideration of spin orbit coupling (SOC) is implemented to describe the electronic structure of TMD systems.

$$\hat{H} = at(\tau k_x \hat{\sigma}_x + k_y \hat{\sigma}_y) + \frac{\Delta}{2}\hat{\sigma}_z - \lambda\tau\frac{\hat{\sigma}_z - 1}{2}\hat{s}_z, \qquad (9)$$

where $\tau = \pm 1$ is the valley index, $\hat{\sigma}$ denotes the Pauli matrices for the two basic functions, $a$ is the lattice constant, $t$ is the hybridization energy (hopping integral), $\Delta$ is the site energy difference, $\hat{s}_z$ is the Pauli matrix for spin, and $2\lambda$ is the spin splitting at the valence band maximum caused by spin orbit coupling (SOC). The berry curvature is then derived from the eigenvalues and eigenfunctions of Eq. (9).

$$\Omega_V(k) = \tau\frac{2a^2t^2\Delta'}{[\Delta'^2 + 4a^2t^2k^2]^{3/2}}, \qquad (10)$$

where $\Delta'(k) \equiv \Delta(k) + \tau s_z\lambda$ is the spin dependent band gap, and $s_z$ is the quantum number $\hbar/2$. The valley splitting ($\mu$) defined as the difference between eigenvalues of the top valence band at inequivalent K points is equal to the spin splitting $2\lambda$ according to the band-structure solved in our TB model. After representation transformation of Eq. (10), the Valley Hall conductivity for TMD can be written as

$$\sigma_{xy}^V = \frac{e^2}{4h}\left(\frac{\Delta'(+K)}{\varepsilon_{F(+K)}} - \frac{\Delta'(-K)}{\varepsilon_{F(-K)}}\right), \qquad (11)$$

where $\varepsilon_{F(\pm K)}$ are the eigenvalues at valleys. Accordingly, the Valley Hall current excited by the strain induced pseudoelectric field ($E$) takes the form as

$$J_V = \frac{Ee^2}{4h}\left(\frac{\Delta + s_z\lambda}{\sqrt{(\Delta + s_z\lambda/2)^2 + 4\pi^2 t^2}} - \frac{\Delta - s_z\lambda}{\sqrt{(\Delta - s_z\lambda/2)^2 + 4\pi^2 t^2}}\right). \qquad (12)$$

It turns out that the Valley Hall current is zero in systems with no spin splitting and thus no valley splitting ($\lambda = \mu = 0$). Therefore, the valley piezoelectricity emerges only in valley polarized systems with spin-valley coupling effect. With the band-structure modified by SOC, the spin splitting at inequivalent valleys exhibits opposite signs, leading to asymmetry distribution of Berry curvature. Considering that $\lambda \ll \Delta$, the Valley Hall current $J_V$ is approximately proportional to $\lambda$ as illustrated in Figure 2a. In addition, $J_V$ decreases with increasing hybridization energy $t$, owing to the larger covalency with rising $t$ that mitigates the charge transport. Moreover, $J_V$ increases with reducing site energy difference $\Delta$, which can be ascribed to the enhanced charge localization with increasing band gap. It is worth noting that while the influence of spin-orbit coupling on Valley Hall current contributed by the top valence band can be qualitatively revealed by the two-band tight-binding model, quantitative prediction on the relation between spin-orbit coupling and valley piezoelectricity requires the incorporation of both nearly degenerated bands with distinct spins to account for the impacts of all valleys. To this end, the Hamiltonian needs to be rewritten as a complicated 4 × 4 matrix, which is beyond the scope of this work.

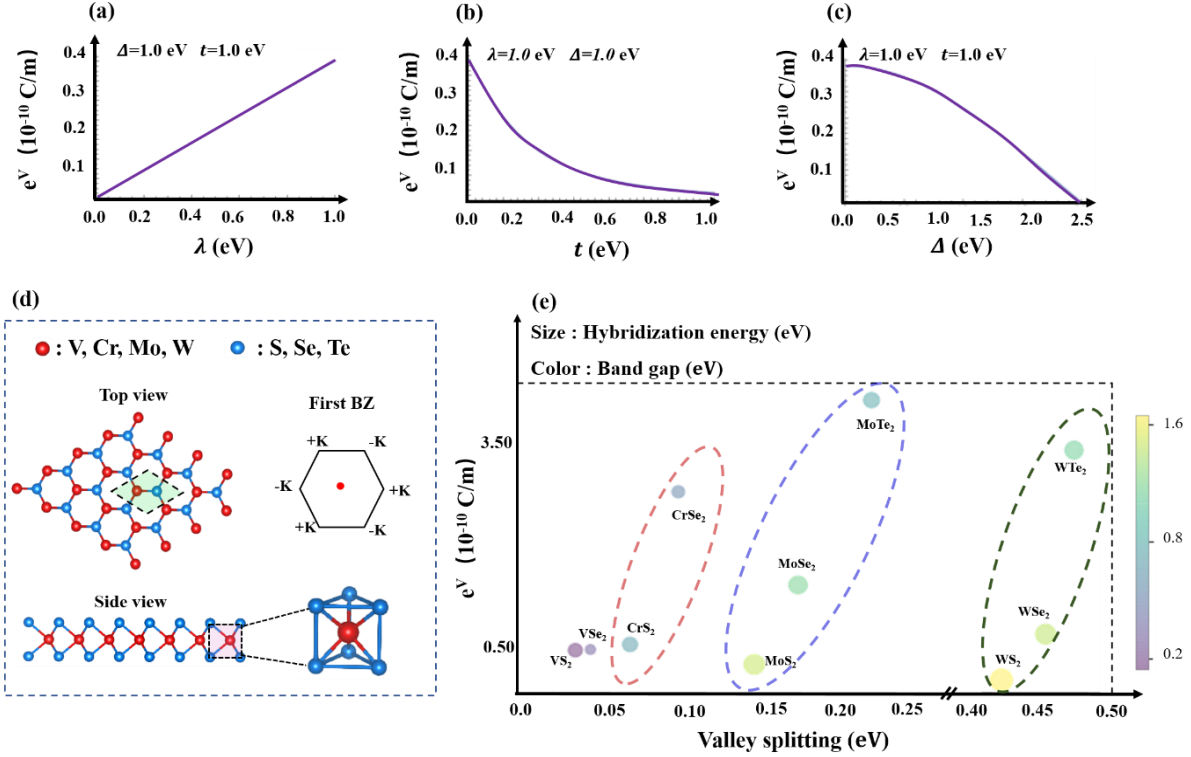

**Figure 2**. Dependence of valley piezoelectricity on fundamental material variables. Correlations between valley current and (a) spin splitting, (b) hybridization energy, (c) site energy difference predicted by analytical model for TMD monolayers. (d) Atomic structure of TMD monolayers considered in DFT calculations. (e) Impact of valley splitting, hybridization energy, and bandgap on valley piezoelectricity unveiled by DFT simulations for TMD monolayers. Hybridization energy and bandgap are illustrated by color and size of circles respectively.

To further justify the trends obtained by the TB analysis, we performed DFT calculations on a variety of TMD systems, with the results summarized in Table 1 and Figure 2e. The hopping energy $t$ is obtained by fitting the DFT band-structure to the TB solution, while other parameters including valley/spin splitting, Poisson ration, and piezoelectric coefficient with/without SOC are directly acquired by DFT simulations. The valley piezoelectricity is computed as the difference between piezoelectric coefficients with and without SOC. As an example, the calculated piezoelectric coefficient of VSe$_2$ monolayer without SOC is $e_{yy} = 4.004$ ($10^{-10}$ C/m), while piezoelectricity tensor with SOC is

$$e_{ijk} = \begin{pmatrix} -0.728 & 0.728 & 0 \\ -4.732 & 4.732 & 0 \\ 0 & 0 & 0 \end{pmatrix}. \tag{13}$$

The $e_{ijk}$ attained by DFT takes the same form as Eq. (8), with the matrix elements satisfying $-e_{xx}^V = -e_{yx}^V = e_{xy}^V = e_{yy}^V$. Such result validates our valley piezoelectric model and suggests that the contribution of valley polarization is too large to be neglected in quantum 2D materials.

**Table 1.** Valley piezoelectricity of TMD monolayers with different variables.

|  | Valley splitting (eV) | Spin splitting (eV) | Bandgap (eV) | Poisson ratio | Hybridization energy (eV) | $e_{yy}$ ($10^{-10}$ C/m) | Valley $e_{xx}$ ($10^{-10}$ C/m) | Valley $e_{yy}$ ($10^{-10}$ C/m) |
| --- | --- | --- | --- | --- | --- | --- | --- | --- |
| **VS$_2$** | 0.034 | 0.958 | 0.151 | 0.259 | 0.889 | 4.153 | - 0.719 | 0.719 |
| **VSe$_2$** | 0.043 | 0.930 | 0.389 | 0.330 | 0.472 | 4.004 | - 0.728 | 0.728 |
| **CrS$_2$** | 0.069 | 0.069 | 0.849 | 0.230 | 1.018 | 4.002 | - 0.791 | 0.790 |
| **CrSe$_2$** | 0.099 | 0.091 | 0.635 | 0.265 | 0.727 | 3.702 | - 2.722 | 2.668 |
| **MoS$_2$** | 0.143 | 0.147 | 1.591 | 0.228 | 1.883 | 3.052 | - 0.543 | 0.544 |
| **MoSe$_2$** | 0.176 | 0.183 | 1.320 | 0.235 | 1.518 | 2.677 | - 1.541 | 1.522 |
| **MoTe$_2$** | 0.223 | 0.214 | 0.922 | 0.269 | 1.126 | 2.401 | 3.796 | 0.558 |
| **WS$_2$** | 0.419 | 0.421 | 1.781 | 0.214 | 2.288 | 2.123 | - 0.355 | 0.363 |
| **WSe$_2$** | 0.454 | 0.457 | 1.562 | 0.190 | 1.867 | 1.873 | - 0.918 | 0.918 |
| **WTe$_2$** | 0.476 | 0.479 | 1.273 | 0.279 | 1.541 | 1.546 | 3.147 | 0.074 |

Since both the valleys in TMD systems and the spin-orbit coupling effects are dominantly contributed by the heavy metal atoms, we compared the valley piezoelectricity of materials with the same metal atoms and possess similar band-structures (grouped by circles) to reveal the influence of different variables on valley piezoelectricity. In general, the hybridization energy and band gap rise with the electronegativity of chalcogenide atoms (S, 6.22; Se, 5.89; Te, 5.49). The trend of XTe$_2$ > XSe$_2$ > XS$_2$ in valley piezoelectricity predicted by DFT calculations can thus be ascribed to the decline of hybridization energy and band gap combined with the enhancement of valley splitting, which is consistent with the trends predicted by analytical model. The valley splitting is no longer equal to the spin splitting in DFT results because the complicated impact of SOC on band-structure can not be sufficiently described by the simplified Hamiltonian in tight binding model. Detailed comparison of valley piezoelectricities between different TMD systems based on band-structure analysis will be provided in the next section.

**2.3. Generality and Transferability of Valley Piezoelectricity**

In order to demonstrate the generality and transferability of our valley piezoelectricity model, we conducted systematic DFT analyses on the electronic structure, berry curvature, and piezoelectric coefficients of a variety of 2D quantum materials including the TMD, group IV monochalcogenides, and transition metal halogenides monolayers as shown in Figure 3. While both the TMD and $Nb_3I_8$ monolayers possess noticeable valley piezoelectricities in x and y directions, the valley piezoelectricity disappears in the group IV monochalcogenides such as SnS monolayer (Figure 3a). According to the valley piezoelectricity model, this can be explained by the valley splitting at high symmetry points +K and -K of TMD and $Nb_3I_8$ monolayers (Figure 3b, d, e), and the absent valley splitting at valleys of group IV monochalcogenides (Figure 3c).

Four representative materials including GeS, GeSe, SnS, and SnSe are considered in group IV monochalcogenides. Similar band-structures are obtained by DFT calculations, with the valleys of top valence bands located at the Y-G lines of reciprocal space. Weak SOC effects are observed as expected from the missing transitional metal and heavy elements in such materials, and the spin splittings at valleys are further eliminated by the crystal symmetry. The valleys around Y and -Y points thus remain equivalent and contribute opposite Berry curvatures, which leads to the vanishing of both valley Hall conductivity and valley piezoelectricity.

The TMD monolayers exhibiting a wide range of valley piezoelectricities can be divided into two categories: the group I composed of $VX_2$ and the group II including $CrX_2$, $MoX_2$, $WX_2$. The group I materials contain odd number of valence electrons, and therefore the pairs of orbitals with up and down spins near the Fermi level serve as top valence band and bottom conduction band respectively (Figure 3e). The similar valley piezoelectricity of $VS_2$ and $VSe_2$ (0.72 and 0.73×$10^{-10}$ C/m) can be explained by their similar valley splittings of 33.7 and 43.4 meV between the +K and -K points (Figure 3e). Isotropic Poisson ratio (Figure 3f) and identical valley piezoelectric responses in x and y directions are observed for $VS_2$ and $VSe_2$. The group II materials contain even number of valence electrons, hence the top valence band is comprised of two approximately degenerate bands associated with different spins. The berry curvature contributed by all valence bands reaches its extremums at K valleys and deviates from antisymmetry distribution due to the SOC effects as exemplified in $MoS_2$ (Figure 2g). The valley piezoelectricity crucially depends on the difference in absolute values of berry curvature at inequivalent valleys. The $MoTe_2$ and $WTe_2$ monolayers exhibit negative valley piezoelectricities as opposed to other TMD materials, which can be ascribed to the shift of absolute maximum points in Berry curvature from the +K to the -K valleys.

The newly reported $Nb_3I_8$ monolayer is studied as another valley material,[18] with its structure shown in Figure S4. The $Nb_3I_8$ monolayer possesses a band-structure (Figure 3d) similar to that of $VSe_2$ monolayer, but a much smaller valley splitting of 5.9 meV. Such weak

SOC effect results in a low valley piezoelectricity of $e_{xx} = -0.1(10^{-10}$ C/m), $e_{yy} = 0.3(10^{-10}$ C/m). Consistent with the anisotropic distributions Poisson ratio (Figure 3f), both the $Nb_3I_8$ and $XTe_2$ monolayers exhibit anisotropic piezoelectric coefficients because of the strong coupling between the metal and anion atoms (Figure 3a). The valley piezoelectricities of $Nb_3I_8$ and TMD monolayers completely vanish if the SOC effects are artificially shunt off, which confirms the critical role of broken time reversal symmetry.

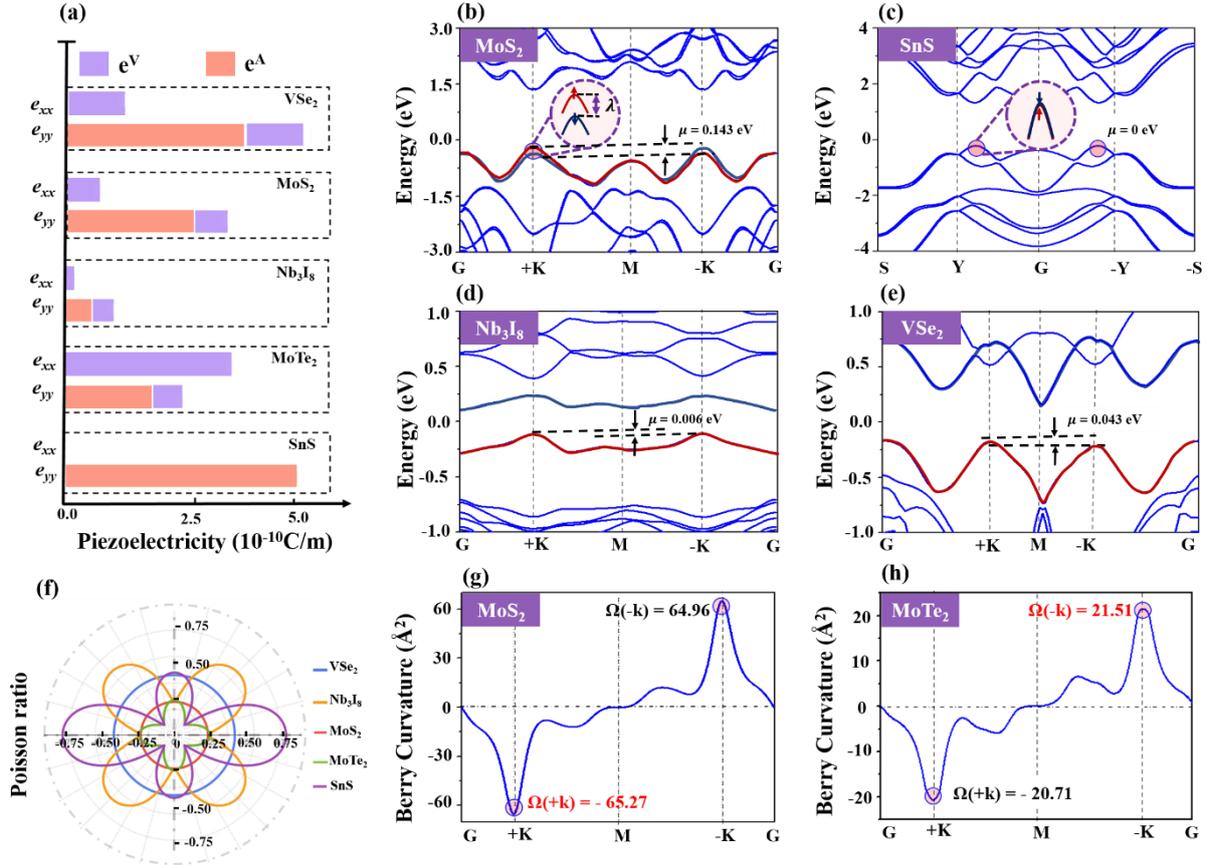

**Figure 3.** Spin-orbit coupling induced valley piezoelectricity for a variety of 2D quantum materials. (a) Traditional and valley piezoelectricities of monolayer materials. Band-structures of (b) $MoS_2$, (c) SnS, (d) $Nb_3I_8$, and (e) $VSe_2$ monolayers. (f) Orientation dependent Poisson ration of valley materials. Berry curvatures of (g) $MoS_2$ and (h) $MoTe_2$ monolayers. The results are obtained by DFT simulations with SOC effect.

## 2.4. Optimization Strategies of Valley Piezoelectricity

Based on the valley piezoelectricity model, manipulating valley splitting is a promising approach to optimize piezoelectricity of quantum materials. The modified $MoS_2$ monolayers were taken as representative examples to illustrate the rational design of band structure towards large piezoelectricity (Figure 4). External stress has been widely implemented to improve piezoelectricity, and we confirmed that a 4% strain leads to a 29.8% enhancement of piezoelectricity. According to previous theoretical study, the reductions in bandgap and hybridization energy of 0.276 eV and 0.25 eV (Figure 4a) would result in a slight augmentation of traditional piezoelectricity. On the other hand, the 18.31 meV enhancement of valley splitting in combination with the reductions of bandgap and hybridization energy would intensify the valley piezoelectricity as suggested by our analytical model. Therefore, the rise of total piezoelectricity originates from the increase of both traditional and valley piezoelectricities.

Doping and surface functionalization may serve as alternative methods to improve piezoelectricity by introducing defect bands with large valley splitting. To justify such speculation, we considered the 5 × 5 supercell of $MoS_2$ monolayer containing Ni, V, and Cr atoms modified by transition metal atoms. Since the SOC simulation for such large supercell is too expensive, the DFT calculations with spin polarization but without SOC were conducted for the passivated and doped systems. With a Cr atom adsorbed on top of the $MoS_2$, the original band-structure of $MoS_2$ is mostly preserved while seven additional defect bands appear within the gap (Figure 4c). The defect bands below the fermi level are dominantly comprised of the Mo and Cr orbitals, with several valley structures emerging at the K points. The piezoelectricity without SOC rises to 5.22, 72.9% larger than the pristine $MoS_2$ monolayer (Figure 4b). The piezoelectric improvement can be ascribed to the enhancement of traditional piezoelectricity contributed by the extra valleys in defect bands as well as the valley piezoelectricity arising from the slightly asymmetric defect band-structure. Considering both Cr and Mo atoms have SOC effects, the actual piezoelectricity should be larger than the calculated value.

Despite the larger dispersion of defect band when the dopant atom is inserted into the crystal rather than located at the surface, the $MoS_2$ monolayer with one Mo atom replaced by either a Cr or a V atom exhibits little promotion on piezoelectricity compared to the pristine $MoS_2$ (Figure 4b). This is because the majority of defect bands are unoccupied states with no contribution to piezoelectricity, and the only occupied defect band is too flat to provide noticeable berry curvature. Similarly, the introduction of a Mo vacancy in the $MoS_2$ supercell has tiny influence on piezoelectricity due to the negligible berry curvature of the flat bands generated by the localized defect states.

To address these issues, the $MoS_2$ monolayer is instead doped by a Ni atom. Seven occupied defect bands composed of Ni, Mo, and S orbitals are observed within the gap, in which several asymmetric valleys appear around the fermi level (Figure 4d). The piezoelectricity without SOC substantially rises to 6.91, 126.6% larger than the pristine $MoS_2$ monolayer (Figure 4b). Similar to the Cr adsorption case, the improvement arises from the increase of both traditional and valley piezoelectricity. Given the strong SOC of Ni atom and the small gap between defect bands, we expect a strong valley piezoelectric response in the Ni doped MoS$_2$ and thus a much large piezoelectricity than the calculated value with SOC. The above analysis suggest that the dopant should be selected to possess large SOC coefficient and suitable energy level alignment with the crystal orbitals for the sake of optimizing piezoelectricity in quantum materials.

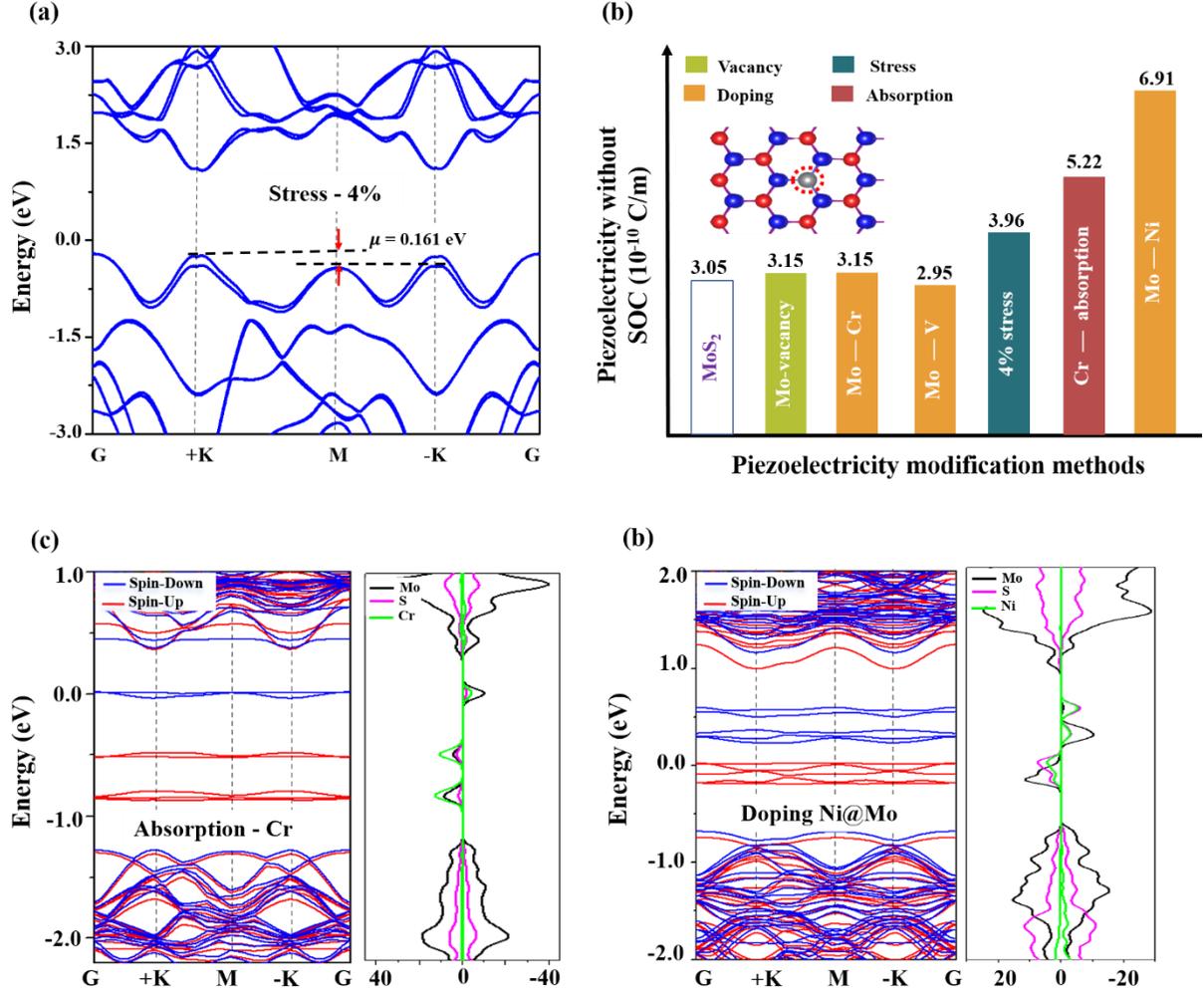

**Figure 4.** Optimization strategies of piezoelectric response in 2D quantum materials. (a) Band-structure of 4% stressed MoS$_2$ monolayer calculated with SOC. (b) Piezoelectricities of modified MoS$_2$ monolayer calculated without SOC, with the doping configuration shown in the inset. Band-structures and partial density of states of (c) Cr absorbed and (d) Ni doped MoS$_2$ monolayers.

## 3. Conclusion

In conclusion, we discovered the electro-mechanical response arising from valley polarization, which is absent in traditional piezoelectric scenario yet crucial for comprehensively describing piezoelectricity in quantum materials. We developed a theoretical model to elucidate the valley piezoelectric mechanism as the activation of Valley Hall conductivity by the pseudoelectric field originating from scalar potential. In contrast to the requirement of broken inversion symmetry for traditional piezoelectricity, broken time reversal symmetry is necessary to destroy the antisymmetry of berry curvature and thus to stimulate non-vanishing valley piezoelectricity. The validity and generality of our model are demonstrated by tight-binding analyses as well as DFT calculations on a variety of 2D quantum materials. The results suggest that valley piezoelectricity is sensitive to valley splitting, hybridization energy, bandgap and Poisson ratio. Accordingly, doping, passivation and external stress are proposed as efficient strategies to optimize valley piezoelectricity. Our valley piezoelectric model bridges the gap between electro-mechanical response and quantum effects including spin-valley and spin-orbit couplings. Such mechanism not only has potential to overcome the bottleneck of the relatively weak piezoelectricity in quantum materials, but also provides a tunable platform for the large family of valley materials to be implemented in sensor, actuator, telecommunication, and medical applications.

## 4. Experimental Methods

Standard ab-initio simulations within the framework of DFT were performed using the Vienna Ab Initio Simulation Package (VASP v5.4).[19] Plane-wave and projector-augmented-wave (PAW) type pseudopotentials were employed with a 450 eV kinetic-energy cutoffs. The GGA-PBE functional[20] was used to describe the exchange-correlation interactions. Both the lattice parameters and atomic coordinates were fully relaxed until all forces were smaller than 0.01 eV/Å. K-point grids of 18×18×1 and 14×14×1 were used for TMDs and group IV monochalcogenides, respectively. Vacuum regions of 15 Å in perpendicular direction were applied to prevent artificial interactions. The piezoelectric tensors were computed by the density-functional perturbation theory (DFPT) with a plane-wave cutoff of 550 eV. The WANNIER90 package[21] is employed for generating maximally-localized Wannier (MLW) functions. For TMD monolayers, the relevant states consist of 14 bands formed by hybridization of metal $d$ orbitals and chalcogen $p$ orbitals. Therefore, only the $p$ and $d$ orbitals were used to construct the MLW. The Berry curvatures were then calculated based on the MLW with the WANNIER90 package. The spin-orbit-coupling (SOC) effects were taken into account in all electronic-structure and piezoelectric calculations except for the large supercell systems.

**Acknowledgements**

The authors acknowledge the support from the National Natural Science Foundation of China (52072417, 11804403, 11832019), the Natural Science Foundation of Guangdong